\documentclass[conference]{IEEEtran}
\IEEEoverridecommandlockouts
\usepackage{graphicx}
\usepackage{cite}
\usepackage{amssymb}
\usepackage{amsfonts}
\usepackage{fancyhdr}  %
\usepackage{extarrows}
\usepackage{algorithm}
\usepackage{multirow,tabularx}
\usepackage{mathtools}
\usepackage{xcolor}
\usepackage{bm}
\usepackage{algorithmic}
\usepackage{amsmath}
\usepackage{makecell}
\usepackage{colortbl}
\usepackage{bbm}
\usepackage{caption}
\usepackage{subcaption}
\usepackage{tikz}
\usepackage{pgfplots}
\pgfplotsset{compat=1.18} 
\usepackage{verbatim}
\usepackage{acronym}
\usepackage[hyperfirst=true]{glossaries}
\usepackage{etoolbox}
\newacronym{ISAC}{ISAC}{integrated sensing and communication}
\newacronym{BS}{BS}{base station}
\newacronym{UE}{UE}{user equipment}
\newacronym{SNR}{SNR}{signal-to-noise ratio}
\newacronym{OFDM}{OFDM}{orthogonal frequency division multiplexing}
\newacronym{MIMO}{MIMO}{multiple-input multiple-output}
\newacronym{GLRT}{GLRT}{generalized likelihood ratio test}
\newacronym{LoS}{LoS}{line-of-sight}
\newacronym{ULA}{ULA}{uniform linear array}
\newacronym{AWGN}{AWGN}{additive white gaussian noise}
\newacronym{RCS}{RCS}{radar cross-section}
\newacronym{BER}{BER}{bit error rate}
\newacronym{SR}{SR}{sensing requirement}
\newacronym{CA-CFAR}{CA-CFAR}{cell averaging constant false alarm rate}
\newacronym{PDR}{PDR}{pilot-to-data ratio}

\usepackage{array} 
\usepackage{diagbox}
\usepackage{balance}
\usepackage{float}
\allowdisplaybreaks

\title{Monostatic ISAC Without Full Buffers: Revisiting Spatial Trade-Offs Under Bursty Traffic}

\author{\IEEEauthorblockN{Mauro Marchese\IEEEauthorrefmark{1}, Musa Furkan Keskin\IEEEauthorrefmark{2}, Pietro Savazzi\IEEEauthorrefmark{1}\IEEEauthorrefmark{3}, Henk Wymeersch\IEEEauthorrefmark{2}} \\
\IEEEauthorblockA{\IEEEauthorrefmark{1}University of Pavia, Italy,  \IEEEauthorrefmark{2}Chalmers University of Technology, Sweden, \\ \IEEEauthorrefmark{3}CNIT Consorzio Nazionale Interuniversitario per le Telecomunicazioni, Pavia, Italy \\
E-mail: mauro.marchese01@universitadipavia.it}
\thanks{This work was partly supported by the SNS JU project 6G-DISAC under the EU’s Horizon Europe research and innovation program under Grant Agreement No 101139130, the Swedish Research Council (VR) through the project 6G-PERCEF under Grant 2024-04390, and the European Union under the Italian National Recovery and Resilience Plan (NRRP) of NextGenerationEU, partnership on “Telecommunications of the Future” (PE00000001 - program "RESTART”).}
}

\begin{document}

\bstctlcite{IEEEexample:BSTcontrol}

\maketitle

\begin{abstract}
This work investigates the spatial trade-offs arising from the design of the transmit beamformer in a monostatic \gls{ISAC} \gls{BS} under bursty traffic, a crucial aspect necessitated by the integration of communication and sensing functionalities in next-generation wireless systems. In this setting, the \gls{BS} does not always have data available for transmission. This study compares different \gls{ISAC} policies and reveals the presence of multiple effects influencing \gls{ISAC} performance: \gls{SNR} boosting of data-aided strategies compared to pilot-based ones, saturation of the probability of detection in data-aided strategies due to the non-full-buffer assumption, and, finally, directional masking of sensing targets due to the relative position between target and user. Simulation results demonstrate varying impact of these effects on \gls{ISAC} trade-offs under different operating conditions, 
thus guiding the design of efficient \gls{ISAC} transmission strategies.
\end{abstract}


\glsresetall

\section{Introduction}

\Gls{ISAC} is gaining momentum as a transformative paradigm for 6G networks. Unlike traditional designs where radar and communication systems evolved separately, \gls{ISAC} seeks to exploit their synergy through shared spectrum, hardware, and signal processing \cite{Liu2020,Liu2023}. Over the years, extensive surveys and application studies have outlined its potential across diverse domains, from automotive perception to mobile networks, while mapping key enablers such as waveform design, interference management, and system integration \cite{Liu2020,Ma2020,Zhang2022,Liu2023}. More recent contributions highlight the opportunities brought by intelligent propagation environments and advanced hardware architectures \cite{Meng2025}, as well as the growing role of artificial intelligence in realizing adaptive and scalable solutions for  wireless \gls{ISAC} systems \cite{Zhang2025}.

A central theme in the \gls{ISAC} literature concerns the fundamental trade-offs between communication and sensing functionalities \cite{keskin2025fundamental,Keskin2021,Du2024,Li2025,Lai2024,Xiong2024}. 
Early investigations analyzed waveform design limitations in \gls{OFDM} dual-functional systems, highlighting delay–Doppler ambiguities and performance compromises \cite{Keskin2021}. Building on this, constellation shaping approaches have been proposed to flexibly reshape the \gls{ISAC} trade-off under \gls{OFDM} signaling \cite{Du2024}, while symbol-level precoding and range–Doppler sidelobe suppression techniques addressed similar challenges in \gls{MIMO}-\gls{OFDM} settings \cite{Li2025}. Other works studied the impact of constellation choices on detection performance from a subspace perspective \cite{Lai2024}. From a theoretical standpoint, the deterministic–random (time–frequency) trade-off has been formalized, leading to the notion of sensing-optimal operating points \cite{Xiong2024}. Most recently, a holistic framework has been proposed to jointly capture deterministic–random and spatial trade-offs in monostatic \gls{ISAC}, offering fundamental insights toward 6G integration \cite{keskin2025fundamental}.

Although the literature presents several works analyzing the various trade-offs in monostatic \gls{ISAC} systems, a common baseline assumption underlies these studies: the \gls{ISAC} transceiver is always assumed to have data available for transmission to the communication receiver, or equivalently, the transmitter buffer is considered to be perpetually full. This assumption simplifies reality, where data packets arrive at the \gls{ISAC} transceiver at a given rate. Hence, in practice traffic is bursty: there are periods when the transmitter’s data buffer is empty or under-utilized \cite{zhou2014toward}. In such cases, a data-only \gls{ISAC} system would simply fall silent, causing sensing to halt whenever there is no user data to send. This issue has only recently been considered in \cite{he2023sencom}, which injects additional packets to maintain a stream of reflections for sensing. Another approach is to schedule dedicated sensing intervals adaptively. For instance, reinforcement learning has been used to dynamically allocate time slots for sensing whenever data queues empty or latency allows \cite{ vaca2023proximal}, effectively borrowing idle communication capacity for radar functions. 
These emerging works clearly demonstrate that to integrate sensing and communication in real networks, the full-buffer assumption must be relaxed, leading to several questions: 
(i) what are the main effects that rule the performance of standard \gls{ISAC} policies (e.g., {pure communication}, {concurrent transmission} and {time sharing}), under the assumption of \textit{bursty traffic at the \gls{BS}}; and (ii)
 how does relaxing the full-buffer assumption at a BS performing monostatic sensing affect the trade-offs arising from the choice of the \gls{ISAC} transmit beamformer?  

In this work, \gls{ISAC} is considered from the perspective of bursty data traffic and the impact of various policies on \gls{ISAC} performance is studied. The specific contributions of this work are: (i) a novel problem formulation of \gls{ISAC} and beamforming strategies under bursty traffic and specific sensing requirements is provided, significantly extending the classical full-buffer models; (ii) a \gls{GLRT}-based method with coherent integration is developed for multi-target detection, considering the existence of a sensing window during which a target should be detected; (iii) a numerical analysis of sensing and communication performance under varying \gls{ISAC} requirements and environmental conditions is performed and guidelines for the design of \gls{ISAC} systems under more realistic traffic conditions are provided.


\section{System Model}
This section details the ISAC model including the bursty traffic model. The sensing and communication requirements are also specified. 

\subsection{ISAC Scenario under Bursty Traffic Paradigm}
The scenario involves a dual-functional, multi-antenna \gls{BS} equipped with both a transmit array and a sensing array, each having $N$ antennas, and an infinite-length buffer for data packets. Moreover, the environment includes $K$ sensing targets, and $U$ single-antenna \glspl{UE}. 
For each user, communication packets arrive at the \gls{BS} following a Poisson process with rate $\lambda_u$ (packets per second) and each packet comprises $B$ bits.

The system operates at carrier frequency $f_c$ and the carrier wavelength is $\lambda_c=c/f_c$, where $c$ denotes the speed of light. Each target is at angle $\tilde\theta_{k}$ and distance $\tilde{d}_k$ and each \gls{UE} at angle $\theta_{u}$ and distance $d_u$. Based on previous angle estimation, the user's angles are assumed to be known at the \gls{BS}, and the \gls{BS} serves one \gls{UE} at a time.
In order to simplify the problem, the \gls{BS} is assumed to transmit symbols using a single-carrier modulation scheme and the \gls{BS}-to-\gls{UE} and \gls{BS}-to-target channels can be considered as \gls{LoS} channels. Moreover, the \gls{BS} uses a unit norm precoder $\mathbf{f}$ to send a symbol $x$ and it is subject to an average power constraint. 
The signal sent by the \gls{BS} in each symbol time is given as
$\mathbf{s}(t)=\sqrt{E_s}\mathbf{f}g(t)x$,
where $E_s$ is the average energy per symbol, $x$ is the transmit symbol (with $\mathbb{E}\big[|x|^{2}\big]=1$) taken from a $Q$-ary modulation alphabet and $g(t)$ is a Nyquist pulse with bandwidth $W$ and unitary energy. Hence, the symbol time is $1/W$. It can be noted that the rates of packet arrivals should satisfy $B\sum_u\lambda_u <W\log_2Q$ in order to have a bursty traffic. Otherwise, the \gls{BS} has always data to send (full buffer case).

The signal received by the $u$-th \gls{UE} is given by
\begin{equation}
r_{u}(t)=\alpha_u\mathbf{a}^\top(\theta_u)\mathbf{s}(t)+n(t),
\end{equation}
where $\alpha_{u}$ is a fixed channel gain obtained using the free space path loss (Friis) equation $|\alpha_u|^2=\lambda_c^2/(4\pi d_u)^2$ and $n(t)$ is \gls{AWGN} noise with spectral density $N_0$. Moreover, $\mathbf{a}(\theta)$ is the steering vector of the \gls{ULA} at the \gls{BS}, defined as $\mathbf{a}(\theta)=[1 \ \ e^{j\pi\sin(\theta)} \ \ \dots \ \ e^{j\pi(N-1)\sin(\theta)}]^\top$, assuming a spacing between the antenna elements of $\lambda_c/2$. After matched filtering operation, the received observation is given by
\begin{equation}\label{eq:userRxObservation}
y_u=\int r_u(t)g(t)\text{d}t=\alpha_{u}\mathbf{a}^{\top}(\theta_{u})\mathbf{s}+n\in\mathbb{C},
\end{equation}
where $\mathbf{s}=\sqrt{E_s}\mathbf{f}x$ is the transmit vector and $n\sim\mathcal{CN}(0,N_{0})$.

Similarly, the backscattered signal at the \gls{BS} is obtained by
\begin{equation}
\mathbf{r}_s(t)=\sum_{k=1}^K\tilde\alpha_{k}\mathbf{a}(\tilde\theta_{k})\mathbf{a}^{\top}(\tilde\theta_{k})\mathbf{s}(t)+\mathbf{n}(t) ,
\end{equation}
where $\tilde\alpha_{k}$ is the target gain obtained by the radar range equation $|\tilde\alpha_k|^2=\sigma_{\text{rcs},k}\lambda_c^2/[(4\pi)^3\tilde{d}_k^4]$ where $\sigma_{\text{rcs},k}$ is the \gls{RCS} of the $k$-th target and $\mathbf{n}(t)$ is \gls{AWGN}.
The matched filter provides as output
\begin{equation} \label{eq_ys}
\mathbf{y}_s=\int\mathbf{r}_s(t)g(t)\text{d}t=\sum_{k=1}^K\tilde\alpha_{k}\mathbf{a}(\tilde\theta_{k})\mathbf{a}^{\top}(\tilde\theta_{k})\mathbf{s}+\mathbf{n}\in\mathbb{C}^N,
\end{equation}
where $\mathbf{n}\sim\mathcal{CN}(0,N_0\mathbf{I}_N)$.

\begin{figure}[t]
\centering
    \includegraphics[width=\columnwidth]{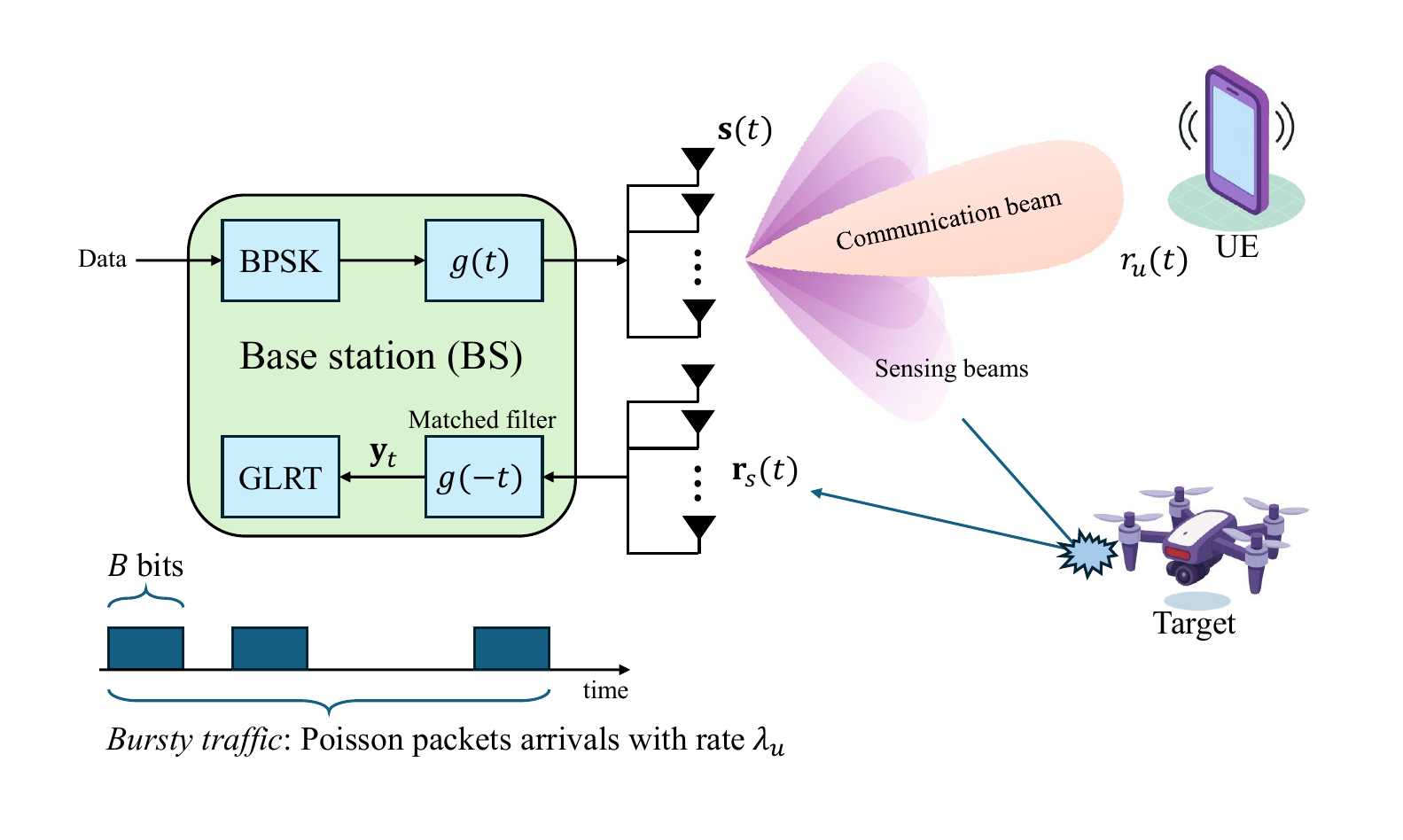}
    \caption{The considered ISAC scenario including a BS performing monostatic sensing, one UE and one target.}\label{fig:FNNarchitecture}
\end{figure}

\subsection{ISAC Requirements}\label{SecISACrequire}
There is a \gls{SR} that a target should be detected every $T_s$ seconds or, equivalently, every $N_s=T_sW$ channel uses. Therefore, the \gls{BS} adopts a policy aiming to maximize the probability of target detection within the sensing window $T_s$. 

For the communication, the policy adopted by the \gls{BS} should also provide the desired communication \gls{SNR} as it directly affects \gls{BER}. Whenever the \gls{BS} transmits a data packet destined to user $u$, the instantaneous communication \gls{SNR} is given as
\begin{equation}
\text{SNR}_{c}=\frac{E_{s}|\alpha_{u}\mathbf{a}^{\top}(\theta_{u})\mathbf{f}|^{2}}{N_{0}}.
\end{equation}
Therefore, the \gls{BER} is minimized when the average value of $|\mathbf{a}^{\top}(\theta_{u})\mathbf{f}|$ over time is maximized.

\section{Precoder Design and Sensing Algorithm}
This section first introduces the \gls{ISAC} strategies considered for the transmit beamformer design, followed by a presentation of the \gls{GLRT}-based multi-target detection algorithm with coherent integration used at the BS.

\subsection{ISAC Policies}\label{secISACstrategies}
The \gls{BS} operates using one of the following \gls{ISAC} strategies:

\subsubsection{Pure Communication}
When data packets are available in the buffer, the \gls{BS} uses the precoder $\mathbf{f}=\mathbf{a}^{*}(\theta_{u})/{\sqrt{N}}$ to steer the beam towards the $u$-th \gls{UE}, which is the current data recipient. In this case, all resources are dedicated to communication, and all the energy is sent towards the \gls{UE}. Therefore, sensing is purely opportunistic. The average power is given by $P_T=\sum_u\lambda_uBE_s/\log_2Q$ by noting that $\sum_u\lambda_uT_s$ is the average number of packets within each sensing window and each packet requires an energy of $BE_s/\log_2Q$ to be transmitted. Finally, the \gls{BS} sets $E_s$ in order to meet the power constraint.

\subsubsection{Time Sharing}
This strategy gives priority to sensing by allocating dedicated time slots for sensing pilots \cite{keskin2025fundamental}.
Specifically, during each sensing window $T_s$, the \gls{BS} allocates $N$ channel uses for sensing and sweeps through a codebook comprising $N$ angular sectors by sending unit amplitude pilots ($x=1$). The codebook spans the angular range $[-\theta_{\max},\theta_{\max}]$ and the $n$-th direction is given as $\theta_n=-\theta_{\max}+2\frac{n-1}{N-1}\theta_{\max}$. During the sweeping process, the \gls{BS} sets $\mathbf{f}=\mathbf{f}_{s}^{(n)}$, where $\mathbf{f}_s^{(n)}=\mathbf{a}^{*}(\theta_{n})/{\sqrt{N}}$. Therefore, 
the transmit vector is then given as $\mathbf{s}=\sqrt{E}_p\mathbf{f}_s^{(n)}$ where $E_p$ denotes the pilot energy. The \gls{ISAC} trade-off is tuned by introducing the \gls{PDR} as $\beta=E_p/E_s\in[0,\infty)$. Here, since the \gls{BS} transmits dedicated pilots for sensing, the average transmit power is given as 
\begin{equation}
P_T=\frac{B\sum_u\lambda_u}{\log_2Q}E_s+\frac{NE_p}{T_s}.
\end{equation} 
Given the \gls{PDR}, the \gls{BS} fixes $E_p$ and $E_s$ in order to meet the average power constraint.
 After the sweeping process is completed, the remaining time $T_s-N/W$ is used for communication and data packets are sent to the desired users by setting $\mathbf{f}=\mathbf{a}^{*}(\theta_{u})/\sqrt{N}$. Hence, it can be noted that the sensing requirement should satisfy $T_s>(B/\log_2Q+N)/W$, otherwise the \gls{BS} doesn't have enough time for sweeping and for sending communication packets.

\subsubsection{Concurrent Transmission}
Whenever a packet arrives, the \gls{BS} uses the precoder \cite{keskin2025fundamental,Pucci2022,Zhang2019}
\begin{equation}
\mathbf{f}=\sqrt{\rho}\mathbf{f}_{s}+\sqrt{\frac{1-\rho}{N}}\mathbf{a}^{*}(\theta_{u}),
\end{equation}
where $\mathbf{f}_{s}=\mathbf{a}^{*}(\theta_{n})/{\sqrt{N}}$ continuously sweeps through the angular codebook for the entire transmission. Moreover, $\rho\in[0,1]$
is an \gls{ISAC} trade-off parameter. A portion of the energy equal to $\rho$ is allocated for sensing and is used for beam sweeping. The remaining portion of energy $1-\rho$ is used for communication to steer a beam towards the desired \gls{UE}. Two extreme cases are identified: when $\rho=0$ all the energy is used for communication and this strategy reduces to \textit{pure communication}; on the other hand, if $\rho=1$, all the energy is used for beam sweeping (\textit{pure sensing}), so, the communication \gls{SNR} is maximum when the \gls{BS} points towards the \gls{UE} and varies periodically. It can be noted that this strategy differs from \textit{time sharing} as the latter uses dedicated pilots for sensing (which eats away resources from communications), while the former leverages random data for sensing.
As for \textit{pure communication}, since this strategy relies solely on data packets, the transmit energy is fixed to $E_s={P_T\log_2Q}/{(B\sum_u\lambda_u)}$ in order to meet the average power constraint.

\subsection{GLRT Detector with Coherent Integration}
During the sensing window $T_s$, the \gls{BS} collects observations depending on the adopted \gls{ISAC} policy. In particular, when the \textit{time sharing} policy is adopted, the \gls{BS} collects $M=N$ observations during the beam sweeping process. Conversely, when \textit{pure communication} or \textit{concurrent transmission} strategies are adopted, the \gls{BS} collects $M$ observations, where $M$ depends on the availability of data for transmission. In this case, $M$ highly depends on the amount of data already present in the buffer and on the time of arrival of a packet, the packet length and the sensing requirement $T_s$. 
\subsubsection{Single-Target Case}
In order to declare the presence of a target, after having collected these $M$ observations $\{\mathbf{y}_s^{(m)}\}_{m=1}^M$ during one sensing window, the \gls{BS} needs to perform hypothesis testing. 
Under hypothesis $H_0$, the target is absent, while under hypothesis $H_1$, the target is present in the environment. Concatenating the $M$ observations, we define
\begin{equation} \label{eq_ybf}
\tilde{\mathbf{y}}=
\begin{bmatrix}
(\mathbf{y}_s^{(1)})^\top\, 
(\mathbf{y}_s^{(2)})^\top \,
\dots \,
(\mathbf{y}_s^{(M)})^\top
\end{bmatrix}^\top\in\mathbb{C}^{MN}.
\end{equation}
Then, considering \eqref{eq_ys}, the detection test is formulated as
\begin{equation}
\tilde{\mathbf{y}}=
    \begin{cases}
    \mathbf{n} & \text{under } H_0, \\
    \mathbf{x}+\mathbf{n} & \text{under } H_1,
    \end{cases}
\end{equation}
where $\mathbf{x} = \tilde\alpha [(\mathbf{a} (\tilde\theta)\mathbf{a}^{\top}(\tilde\theta)\mathbf{s}_1)^\top \dots (\mathbf{a} (\tilde\theta)\mathbf{a}^{\top}(\tilde\theta)\mathbf{s}_M)^\top]^\top$, $\mathbf{s}_m$ is the $m$-th transmit vector, $\mathbf{n}$ is the concatenated noise vector over $M$ observations defined similar to \eqref{eq_ybf}.
The \gls{GLRT} is 
\begin{equation}
\Lambda(\tilde{\mathbf{y}}) = \frac{ \max_{\tilde\alpha, \tilde\theta} p(\tilde{\mathbf{y}} \mid H_1, \tilde\alpha, \tilde\theta)}{ p(\tilde{\mathbf{y}} \mid H_0)} \underset{H_0}{\overset{H_1}{\gtrless}} \eta \ ,
\end{equation}
where $\eta$ is a threshold, $p(\tilde{\mathbf{y}}\mid H_1, \tilde\alpha, \tilde\theta)$ is the probability density function for $\tilde{\mathbf{y}}$ when a target with parameters $\tilde\alpha,\tilde\theta$ is present, and, finally, $p(\tilde{\mathbf{y}} \mid H_0)$ is the probability density function for $\tilde{\mathbf{y}}$ when the target is absent.
Since noise is gaussian, taking the logarithm one gets
\begin{equation}
\|\tilde{\mathbf{y}}\|^2-\min_{\tilde\alpha,\tilde\theta}\sum_{m=1}^M \left\| \mathbf{y}_s^{(m)} - \tilde\alpha \mathbf{a}(\tilde\theta)\mathbf{a}^\top(\tilde\theta)\mathbf{s}_m \right\|^2\underset{H_0}{\overset{H_1}{\gtrless}} \tilde{\eta},
\end{equation}
where $\tilde{\eta}=N_0\log(\eta)$
For a given $\tilde\theta$, the optimum target gain is given by
\begin{equation}
\hat{\tilde\alpha} = \frac{ \sum_{m=1}^M (\mathbf{a}(\tilde\theta)\mathbf{a}^\top(\tilde\theta)\mathbf{s}_m)^\mathsf{H} \mathbf{y}_s^{(m)} }{ \sum_{m=1}^M \left\| \mathbf{a}(\tilde\theta)\mathbf{a}^\top(\tilde\theta)\mathbf{s}_m \right\|^2 }.
\end{equation}
Therefore, the final \gls{GLRT} for target detection with coherent integration is 
\begin{equation}\label{eq:finalGLRT}
\max_{\tilde\theta} \underbrace{\frac{ \left| \sum_{m=1}^M \big(\mathbf{y}_s^{(m)}\big)^\mathsf{H} \mathbf{a}(\tilde\theta)\mathbf{a}^\top(\tilde\theta)\mathbf{s}_m \right|^2 }{ \sum_{m=1}^M \left\| \mathbf{a}(\tilde\theta)\mathbf{a}^\top(\tilde\theta)\mathbf{s}_m \right\|^2}}_{\text{\normalsize $\chi(\tilde\theta)$}} \underset{H_0}{\overset{H_1}{\gtrless}} \tilde{\eta}.
\end{equation}
\vspace{-1em}

\begin{figure*}[t]
    \centering 
    \begin{subfigure}[t]{0.32\textwidth}
        \centering
        \resizebox{\columnwidth}{!}{\definecolor{mycolor1}{rgb}{0.06600,0.44300,0.74500}%
\definecolor{mycolor2}{rgb}{0.86600,0.32900,0.00000}%
\definecolor{mycolor3}{rgb}{0.00000,1.00000,1.00000}%
\definecolor{mycolor4}{rgb}{0.12941,0.12941,0.12941}%
\begin{tikzpicture}

\begin{axis}[%
width=4.521in,
height=3.566in,
scale only axis,
xmin=-50,
xmax=10,
xlabel style={font=\color{mycolor4}},
xlabel={\Large Target RCS [dBsm]},
ymin=0,
ymax=1,
ylabel style={font=\color{mycolor4}},
ylabel={\Large Probability of Detection},
axis background/.style={fill=white},
clip=false,
xmajorgrids,
ymajorgrids,
legend style={legend cell align=left, align=left,legend pos=north west, font=\Large},
ticklabel style={font=\Large}
]

\addplot [color=mycolor1, line width=2.5pt]
table[row sep=crcr]{%
-50	0.00336\\
-47.5	0.008456\\
-45	0.0262\\
-42.5	0.072656\\
-40	0.155296\\
-37.5	0.240184\\
-35	0.289224\\
-32.5	0.309664\\
-30	0.3204\\
-27.5	0.326032\\
-25	0.329184\\
-22.5	0.332944\\
-20	0.331888\\
-17.5	0.331744\\
-15	0.328304\\
-12.5	0.332728\\
-10	0.331336\\
-7.5	0.33472\\
-5	0.332256\\
-2.5	0.333512\\
0	0.33144\\
2.5	0.334432\\
5	0.335472\\
7.5	0.333464\\
10	0.335168\\
};
\addlegendentry{Pure Communication}

\addplot [color=mycolor2, line width=2.5pt,dashed]
 table[row sep=crcr]{%
-50	0.003968\\
-47.5	0.004112\\
-45	0.004232\\
-42.5	0.003984\\
-40	0.003952\\
-37.5	0.004024\\
-35	0.004104\\
-32.5	0.004296\\
-30	0.004256\\
-27.5	0.005\\
-25	0.006552\\
-22.5	0.010896\\
-20	0.028104\\
-17.5	0.086384\\
-15	0.275976\\
-12.5	0.64868\\
-10	0.9436\\
-7.5	0.998616\\
-5	1\\
-2.5	1\\
0	1\\
2.5	1\\
5	1\\
7.5	1\\
10	1\\
};
\addlegendentry{Time Sharing $\beta=1$}

\addplot [color=mycolor2, line width=2.5pt]
  table[row sep=crcr]{%
-50	0.0045\\
-47.5	0.0035\\
-45	0.004\\
-42.5	0.0047\\
-40	0.00433333333333333\\
-37.5	0.00603333333333334\\
-35	0.0114\\
-32.5	0.0246333333333333\\
-30	0.0804\\
-27.5	0.251033333333333\\
-25	0.6223\\
-22.5	0.9335\\
-20	0.998433333333333\\
-17.5	1\\
-15	1\\
-12.5	1\\
-10	1\\
-7.5	1\\
-5	1\\
-2.5	1\\
0	1\\
2.5	1\\
5	1\\
7.5	1\\
10	1\\
};
\addlegendentry{Time Sharing $\beta=200$}

\addplot [color=blue, line width=2.5pt]
 table[row sep=crcr]{%
-50	0.0026\\
-47.5	0.006464\\
-45	0.017992\\
-42.5	0.055544\\
-40	0.13064\\
-37.5	0.225848\\
-35	0.281072\\
-32.5	0.30696\\
-30	0.318304\\
-27.5	0.324304\\
-25	0.329048\\
-22.5	0.333344\\
-20	0.330696\\
-17.5	0.33156\\
-15	0.332176\\
-12.5	0.331904\\
-10	0.330592\\
-7.5	0.334704\\
-5	0.329928\\
-2.5	0.333536\\
0	0.332808\\
2.5	0.336824\\
5	0.332928\\
7.5	0.335392\\
10	0.332984\\
};
\addlegendentry{$\text{Concurrent }\rho\text{ = 0.4}$}

\addplot [color=green, line width=2.5pt]
 table[row sep=crcr]{%
-50	0.001656\\
-47.5	0.002584\\
-45	0.004832\\
-42.5	0.0144\\
-40	0.04132\\
-37.5	0.107416\\
-35	0.205672\\
-32.5	0.273512\\
-30	0.300392\\
-27.5	0.317224\\
-25	0.324592\\
-22.5	0.327776\\
-20	0.327552\\
-17.5	0.330264\\
-15	0.331056\\
-12.5	0.329568\\
-10	0.332056\\
-7.5	0.331416\\
-5	0.332376\\
-2.5	0.335496\\
0	0.336152\\
2.5	0.330696\\
5	0.333456\\
7.5	0.332296\\
10	0.33072\\
};
\addlegendentry{$\text{Concurrent }\rho\text{ = 0.8}$}

\addplot [color=red, line width=2.5pt]
 table[row sep=crcr]{%
-50	0.0014\\
-47.5	0.001768\\
-45	0.00192\\
-42.5	0.002152\\
-40	0.004392\\
-37.5	0.012328\\
-35	0.036208\\
-32.5	0.096856\\
-30	0.192536\\
-27.5	0.266416\\
-25	0.298248\\
-22.5	0.314192\\
-20	0.319584\\
-17.5	0.328216\\
-15	0.329968\\
-12.5	0.332344\\
-10	0.327416\\
-7.5	0.332344\\
-5	0.331344\\
-2.5	0.33164\\
0	0.33232\\
2.5	0.333728\\
5	0.333192\\
7.5	0.330424\\
10	0.33116\\
};
\addlegendentry{$\text{Concurrent }\rho\text{ = 1}$}

\addplot [black, line width=2.5pt, dashed] coordinates {(-50, 0.3) (10, 0.3)};
\addlegendentry{$\lambda_uT_s$}

\draw[<->, ultra thick, black] (axis cs:-40,0.12) -- (axis cs:-17.5,0.12) node [midway, above, sloped, font=\Large] {C-SNR-Boost};
\draw[<->, ultra thick, black] (axis cs:-5,0.35) -- (axis cs:-5,0.98) node [midway, right, font=\Large] {NFB-Loss};

\end{axis}

\end{tikzpicture}%
}
        \caption{Low-$\Delta\theta$, $\tilde\theta_k=43^\circ$}\label{fig:PdRCSsmallSRsmalltheta}
    \end{subfigure}
    \hfill 
    \begin{subfigure}[t]{0.32\textwidth}
        \centering
        \resizebox{\columnwidth}{!}{\definecolor{mycolor1}{rgb}{0.06600,0.44300,0.74500}%
\definecolor{mycolor2}{rgb}{0.86600,0.32900,0.00000}%
\definecolor{mycolor3}{rgb}{0.00000,1.00000,1.00000}%
\definecolor{mycolor4}{rgb}{0.12941,0.12941,0.12941}%
\begin{tikzpicture}

\begin{axis}[%
width=4.521in,
height=3.566in,
scale only axis,
xmin=-50,
xmax=10,
xlabel style={font=\color{mycolor4}},
xlabel={\Large Target RCS [dBsm]},
ymin=0,
ymax=1,
ylabel style={font=\color{mycolor4}},
ylabel={\Large Probability of Detection},
axis background/.style={fill=white},
clip=false,
xmajorgrids,
ymajorgrids,
legend style={legend cell align=left, align=left,legend pos=north west, font=\Large},
ticklabel style={font=\Large}
]

\addplot [color=mycolor1, line width=2.5pt]
  table[row sep=crcr]{%
-50	0.000616\\
-47.5	0.000672\\
-45	0.000656\\
-42.5	0.000672\\
-40	0.000656\\
-37.5	0.000608\\
-35	0.000928\\
-32.5	0.00084\\
-30	0.00112\\
-27.5	0.001768\\
-25	0.005784\\
-22.5	0.019464\\
-20	0.060112\\
-17.5	0.145592\\
-15	0.235856\\
-12.5	0.286528\\
-10	0.306648\\
-7.5	0.321016\\
-5	0.324624\\
-2.5	0.32948\\
0	0.328976\\
2.5	0.332944\\
5	0.334768\\
7.5	0.333088\\
10	0.334976\\
};
\addlegendentry{Pure Communication}

\addplot [color=mycolor2, line width=2.5pt,dashed]
  table[row sep=crcr]{%
-50	0.003944\\
-47.5	0.00412\\
-45	0.004272\\
-42.5	0.003912\\
-40	0.003992\\
-37.5	0.003904\\
-35	0.003968\\
-32.5	0.004216\\
-30	0.00392\\
-27.5	0.00412\\
-25	0.004304\\
-22.5	0.005024\\
-20	0.009536\\
-17.5	0.030688\\
-15	0.114304\\
-12.5	0.383184\\
-10	0.821904\\
-7.5	0.993432\\
-5	1\\
-2.5	1\\
0	1\\
2.5	1\\
5	1\\
7.5	1\\
10	1\\
};
\addlegendentry{Time Sharing $\beta=1$}

\addplot [color=mycolor2, line width=2.5pt]
  table[row sep=crcr]{%
-50	0.0045\\
-47.5	0.00356666666666667\\
-45	0.00386666666666667\\
-42.5	0.00363333333333334\\
-40	0.00406666666666667\\
-37.5	0.00413333333333334\\
-35	0.00496666666666667\\
-32.5	0.00843333333333334\\
-30	0.0270333333333334\\
-27.5	0.105233333333333\\
-25	0.361333333333333\\
-22.5	0.7956\\
-20	0.9918\\
-17.5	1\\
-15	1\\
-12.5	1\\
-10	1\\
-7.5	1\\
-5	1\\
-2.5	1\\
0	1\\
2.5	1\\
5	1\\
7.5	1\\
10	1\\
};
\addlegendentry{Time Sharing $\beta=200$}

\addplot [color=blue, line width=2.5pt]
  table[row sep=crcr]{%
-50	0.001416\\
-47.5	0.001256\\
-45	0.001248\\
-42.5	0.001528\\
-40	0.001488\\
-37.5	0.002112\\
-35	0.004344\\
-32.5	0.012584\\
-30	0.041648\\
-27.5	0.114232\\
-25	0.217832\\
-22.5	0.282336\\
-20	0.303096\\
-17.5	0.316352\\
-15	0.32404\\
-12.5	0.32728\\
-10	0.327664\\
-7.5	0.333184\\
-5	0.329104\\
-2.5	0.332544\\
0	0.33196\\
2.5	0.336128\\
5	0.332472\\
7.5	0.33508\\
10	0.332816\\
};
\addlegendentry{$\text{Concurrent }\rho\text{ = 0.4}$}

\addplot [color=green, line width=2.5pt]
  table[row sep=crcr]{%
-50	0.00136\\
-47.5	0.001448\\
-45	0.001256\\
-42.5	0.001648\\
-40	0.001792\\
-37.5	0.003432\\
-35	0.010776\\
-32.5	0.0386\\
-30	0.105912\\
-27.5	0.210992\\
-25	0.277504\\
-22.5	0.302392\\
-20	0.313456\\
-17.5	0.32244\\
-15	0.32664\\
-12.5	0.327208\\
-10	0.330864\\
-7.5	0.33064\\
-5	0.331608\\
-2.5	0.334624\\
0	0.335256\\
2.5	0.33008\\
5	0.332968\\
7.5	0.33196\\
10	0.330512\\
};
\addlegendentry{$\text{Concurrent }\rho\text{ = 0.8}$}

\addplot [color=red, line width=2.5pt]
  table[row sep=crcr]{%
-50	0.001352\\
-47.5	0.001408\\
-45	0.0014\\
-42.5	0.001392\\
-40	0.001944\\
-37.5	0.004528\\
-35	0.01452\\
-32.5	0.048568\\
-30	0.13044\\
-27.5	0.23164\\
-25	0.283704\\
-22.5	0.306528\\
-20	0.315528\\
-17.5	0.32572\\
-15	0.328608\\
-12.5	0.33144\\
-10	0.327192\\
-7.5	0.332424\\
-5	0.331304\\
-2.5	0.331568\\
0	0.33216\\
2.5	0.333656\\
5	0.332976\\
7.5	0.330224\\
10	0.330968\\
};
\addlegendentry{$\text{Concurrent }\rho\text{ = 1}$}

\addplot [black, line width=2.5pt, dashed] coordinates {(-50, 0.3) (10, 0.3)};
\addlegendentry{$\lambda_uT_s$}

\end{axis}

\end{tikzpicture}%
}
        \caption{Mid-$\Delta\theta$, $\tilde\theta_k=-25^\circ$}\label{fig:PdRCSmidSRmidtheta}
    \end{subfigure}
    \hfill 
    \begin{subfigure}[t]{0.32\textwidth}
        \centering
        \resizebox{\columnwidth}{!}{\definecolor{mycolor1}{rgb}{0.06600,0.44300,0.74500}%
\definecolor{mycolor2}{rgb}{0.86600,0.32900,0.00000}%
\definecolor{mycolor3}{rgb}{0.00000,1.00000,1.00000}%
\definecolor{mycolor4}{rgb}{0.12941,0.12941,0.12941}%
\begin{tikzpicture}

\begin{axis}[%
width=4.521in,
height=3.566in,
scale only axis,
xmin=-50,
xmax=10,
xlabel style={font=\color{mycolor4}},
xlabel={\Large Target RCS [dBsm]},
ymin=0,
ymax=1,
ylabel style={font=\color{mycolor4}},
ylabel={\Large Probability of Detection},
axis background/.style={fill=white},
clip=false,
xmajorgrids,
ymajorgrids,
legend style={legend cell align=left, align=left,legend pos=north west, font=\Large},
ticklabel style={font=\Large}
]

\addplot [color=mycolor1, line width=2.5pt]
  table[row sep=crcr]{%
-50	0.000656\\
-47.5	0.000656\\
-45	0.000664\\
-42.5	0.000712\\
-40	0.000648\\
-37.5	0.000624\\
-35	0.000944\\
-32.5	0.00072\\
-30	0.000792\\
-27.5	0.000784\\
-25	0.000776\\
-22.5	0.000568\\
-20	0.00052\\
-17.5	0.000712\\
-15	0.000824\\
-12.5	0.000896\\
-10	0.000904\\
-7.5	0.001032\\
-5	0.001232\\
-2.5	0.001112\\
0	0.00088\\
2.5	0.000664\\
5	0.000392\\
7.5	0.0002\\
10	8e-05\\
};
\addlegendentry{Pure Communication}

\addplot [color=mycolor2, line width=2.5pt,dashed]
  table[row sep=crcr]{%
-50	0.003896\\
-47.5	0.004032\\
-45	0.004336\\
-42.5	0.004008\\
-40	0.00396\\
-37.5	0.004096\\
-35	0.004296\\
-32.5	0.004488\\
-30	0.004792\\
-27.5	0.006016\\
-25	0.008224\\
-22.5	0.014072\\
-20	0.029392\\
-17.5	0.07524\\
-15	0.184656\\
-12.5	0.389992\\
-10	0.644632\\
-7.5	0.850624\\
-5	0.96\\
-2.5	0.994992\\
0	0.999912\\
2.5	1\\
5	1\\
7.5	1\\
10	1\\
};
\addlegendentry{Time Sharing $\beta=1$}

\addplot [color=mycolor2, line width=2.5pt]
  table[row sep=crcr]{%
-50	0.00426666666666667\\
-47.5	0.00366666666666667\\
-45	0.00453333333333334\\
-42.5	0.0043\\
-40	0.0062\\
-37.5	0.00763333333333334\\
-35	0.0125333333333333\\
-32.5	0.0275666666666667\\
-30	0.0707\\
-27.5	0.1804\\
-25	0.374733333333333\\
-22.5	0.629533333333333\\
-20	0.840366666666666\\
-17.5	0.955533333333333\\
-15	0.995533333333333\\
-12.5	0.999966666666667\\
-10	1\\
-7.5	1\\
-5	1\\
-2.5	1\\
0	1\\
2.5	1\\
5	1\\
7.5	1\\
10	1\\
};
\addlegendentry{Time Sharing $\beta=200$}

\addplot [color=blue, line width=2.5pt]
  table[row sep=crcr]{%
-50	0.00148\\
-47.5	0.001448\\
-45	0.001584\\
-42.5	0.001992\\
-40	0.002408\\
-37.5	0.003528\\
-35	0.007504\\
-32.5	0.017128\\
-30	0.0402\\
-27.5	0.086008\\
-25	0.156688\\
-22.5	0.231296\\
-20	0.27768\\
-17.5	0.30424\\
-15	0.317648\\
-12.5	0.323784\\
-10	0.325704\\
-7.5	0.332008\\
-5	0.328472\\
-2.5	0.332352\\
0	0.331848\\
2.5	0.335744\\
5	0.331976\\
7.5	0.334432\\
10	0.332104\\
};
\addlegendentry{$\text{Concurrent }\rho\text{ = 0.4}$}

\addplot [color=green, line width=2.5pt]
  table[row sep=crcr]{%
-50	0.00144\\
-47.5	0.001632\\
-45	0.00184\\
-42.5	0.0026\\
-40	0.004128\\
-37.5	0.008424\\
-35	0.019488\\
-32.5	0.04852\\
-30	0.09896\\
-27.5	0.171512\\
-25	0.2382\\
-22.5	0.283264\\
-20	0.30468\\
-17.5	0.317656\\
-15	0.324288\\
-12.5	0.325944\\
-10	0.3298\\
-7.5	0.33012\\
-5	0.331392\\
-2.5	0.334496\\
0	0.335176\\
2.5	0.329992\\
5	0.332608\\
7.5	0.331576\\
10	0.330112\\
};
\addlegendentry{$\text{Concurrent }\rho\text{ = 0.8}$}

\addplot [color=red, line width=2.5pt]
  table[row sep=crcr]{%
-50	0.001472\\
-47.5	0.001816\\
-45	0.002064\\
-42.5	0.002712\\
-40	0.004856\\
-37.5	0.011256\\
-35	0.028088\\
-32.5	0.063864\\
-30	0.122536\\
-27.5	0.192984\\
-25	0.252552\\
-22.5	0.29\\
-20	0.308616\\
-17.5	0.321864\\
-15	0.326352\\
-12.5	0.330328\\
-10	0.326264\\
-7.5	0.332064\\
-5	0.331232\\
-2.5	0.331552\\
0	0.332248\\
2.5	0.333696\\
5	0.333\\
7.5	0.330184\\
10	0.330792\\
};
\addlegendentry{$\text{Concurrent }\rho\text{ = 1}$}

\addplot [black, line width=2.5pt, dashed] coordinates {(-50, 0.3) (10, 0.3)};
\addlegendentry{$\lambda_uT_s$}

\draw[<->, ultra thick, black] (axis cs:-7,0.02) -- (axis cs:-7,0.32) node [midway, right, font=\Large] {DM-Loss};

\end{axis}

\end{tikzpicture}%
}
        \caption{High-$\Delta\theta$, $\tilde\theta_k=-58^\circ$}\label{fig:PdRCSbigSRbigtheta}
    \end{subfigure}
    
    \caption{Sensing performance of ISAC policies against target RCS in the strict sensing requirement scenario ($T_s=0.3$ ms).}\label{fig:PdRCS_strictSR}
\end{figure*}
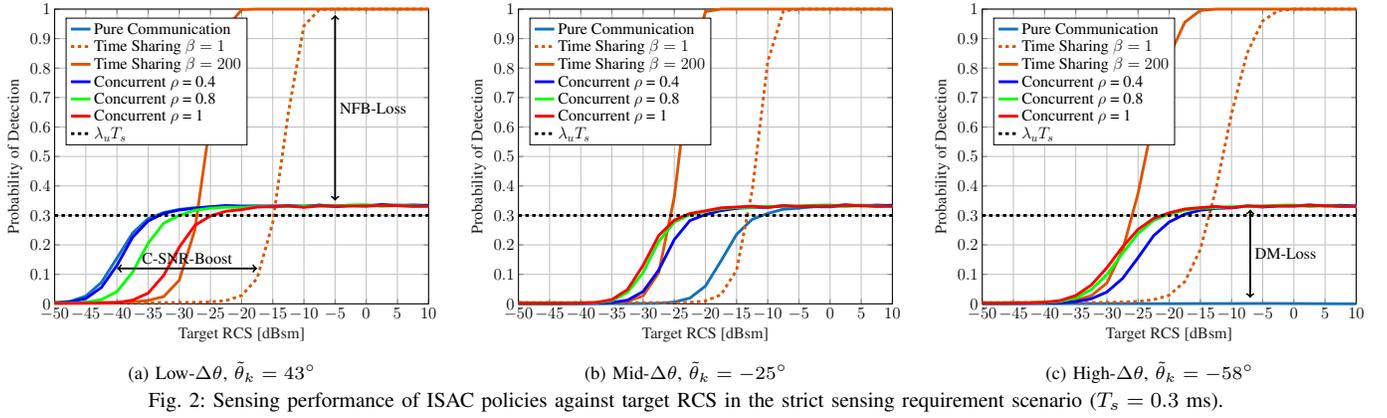

\begin{table}[ht!]
\centering
\caption{Simulation parameters}\label{TabSim}
{
\begin{tabular}{||l | l||}

\hline
Carrier frequency, $f_c$ & $5$ GHz \\
\hline
Transmit power, $P_T$ & $20$ dBm \\
\hline
Number of antennas, $N$ & $16$ \\
\hline
Bandwidth, $W$ & $10$ MHz \\
\hline
Noise power spectral density, $N_0$ & $-174$ dBm/Hz \\
\hline
Packet length, $B$ & $1000$ bits \\
\hline
Packet rate, $\lambda_u/W$ & $10^{-4}$ \\
\hline
Modulation, $Q$ & BPSK, 1 bit/symbol \\
\hline
Distance between UE and BS, $d_u$ & $500$ m \\
\hline
UE direction, $\theta_u$ & $40^\circ$ \\
\hline
Distance between target and BS, $\tilde{d}_k$ & $80$ m \\
\hline
Target direction, $\tilde\theta_k$ & $[43^\circ,-25^\circ,-58^\circ]$ \\
\hline
Maximum sweep angle, $\theta_{\max}$ & $70^\circ$ \\
\hline
\end{tabular}\label{tableSimParam}
}
\end{table}

\subsubsection{Multi-Target Case}
To identify the presence of multiple targets, the
decision statistic $\chi(\tilde\theta)$ in \eqref{eq:finalGLRT} is computed over a
discretized set of angles, and detections are declared at every peak that exceeds a predefined threshold.
A \gls{CA-CFAR} is run to set an adaptive threshold for the \gls{GLRT}. In particular, the cell-averaging is performed as $\tilde{\eta}=(P_{\text{fa}}^{-1/N_c}-1)\sum_{i=1}^{N_c}\chi(\tilde\theta_i)$, where $N_c$ is the number of training cells and $P_{\text{fa}}$ is the desired probability of false alarm.

\section{Simulation Results}
A detailed description of the scenario and its parameters is provided in this section. The results include the sensing performance under various sensing requirements as well as the ISAC performance trade-offs.
\subsection{Simulation Scenario}\label{secSimScenario}
For the sake of simplicity while investigating spatial trade-offs in monostatic \gls{ISAC} under bursty traffic, a scenario including $K=1$ sensing target and $U=1$ user is considered. To gain insight into the impact of geometry, the target is located at an angular separation of $\Delta\theta = |\theta_t - \theta_u|$ from the user. The performance of the different \gls{ISAC} strategies described in Section \ref{secISACstrategies} is investigated under three case studies:
\begin{itemize}
    \item Low-$\Delta\theta$: the target and the \gls{UE} are located in similar directions, falling within the same angular sector;
    \item Mid-$\Delta\theta$: the angular separation between the target and the user is moderate; the target can still be detected when the \gls{BS} steers the beam towards the \gls{UE} due to the beampattern sidelobes;
    \item High-$\Delta\theta$: the angular separation between the target and the user is large enough that the target can no longer be detected when the \gls{BS} steers the beam towards the \gls{UE}.
\end{itemize}
The simulation parameters are summarized in Table \ref{tableSimParam}.

\subsection{Sensing Performance under Bursty Traffic}\label{sec:PdvsRCSandEffects}
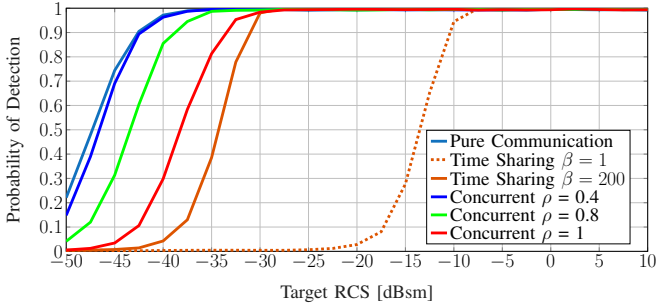
\begin{figure}[t]
    \centering
    \resizebox{0.99\columnwidth}{!}{      \definecolor{mycolor1}{rgb}{0.06600,0.44300,0.74500}%
\definecolor{mycolor2}{rgb}{0.86600,0.32900,0.00000}%
\definecolor{mycolor3}{rgb}{0.00000,1.00000,1.00000}%
\definecolor{mycolor4}{rgb}{0.12941,0.12941,0.12941}%
\begin{tikzpicture}

\begin{axis}[%
width=8.521in,
height=3.566in,
scale only axis,
xmin=-50,
xmax=10,
xlabel style={font=\color{mycolor4}, yshift=-1em},
xlabel={\huge Target RCS [dBsm]},
ymin=0,
ymax=1,
ylabel style={font=\color{mycolor4}, yshift=1em},
ylabel={\huge Probability of Detection},
axis background/.style={fill=white},
xmajorgrids,
ymajorgrids,
legend style={legend cell align=left, align=left,legend pos=south east, font=\huge},
ticklabel style={font=\huge}
]

\addplot [color=mycolor1, line width=3pt]
  table[row sep=crcr]{%
-50	0.220666666666667\\
-47.5	0.479333333333333\\
-45	0.744\\
-42.5	0.904666666666667\\
-40	0.971666666666666\\
-37.5	0.989333333333333\\
-35	0.993666666666666\\
-32.5	0.994\\
-30	0.993\\
-27.5	0.994333333333333\\
-25	0.991666666666666\\
-22.5	0.993666666666666\\
-20	0.992666666666666\\
-17.5	0.994\\
-15	0.993333333333333\\
-12.5	0.993333333333333\\
-10	0.994\\
-7.5	0.993666666666666\\
-5	0.995\\
-2.5	0.992333333333333\\
0	0.995666666666666\\
2.5	0.994333333333333\\
5	0.993333333333333\\
7.5	0.994\\
10	0.992\\
};
\addlegendentry{Pure Communication}

\addplot [color=mycolor2, line width=3pt,dashed]
  table[row sep=crcr]{%
-50	0.00333333333333333\\
-47.5	0.00566666666666667\\
-45	0.003\\
-42.5	0.00266666666666667\\
-40	0.00366666666666667\\
-37.5	0.00433333333333333\\
-35	0.00466666666666667\\
-32.5	0.00466666666666667\\
-30	0.00433333333333333\\
-27.5	0.00433333333333333\\
-25	0.00666666666666667\\
-22.5	0.012\\
-20	0.028\\
-17.5	0.0806666666666667\\
-15	0.277\\
-12.5	0.652666666666667\\
-10	0.944666666666667\\
-7.5	0.999333333333333\\
-5	1\\
-2.5	1\\
0	1\\
2.5	1\\
5	1\\
7.5	1\\
10	1\\
};
\addlegendentry{Time Sharing $\beta=1$}

\addplot [color=mycolor2, line width=3pt]
  table[row sep=crcr]{%
-50	0.00476666666666667\\
-47.5	0.0049\\
-45	0.00766666666666667\\
-42.5	0.0141333333333333\\
-40	0.0426666666666667\\
-37.5	0.129933333333333\\
-35	0.385666666666667\\
-32.5	0.7786\\
-30	0.980466666666667\\
-27.5	0.999766666666667\\
-25	1\\
-22.5	1\\
-20	1\\
-17.5	1\\
-15	1\\
-12.5	1\\
-10	1\\
-7.5	1\\
-5	1\\
-2.5	1\\
0	1\\
2.5	1\\
5	1\\
7.5	1\\
10	1\\
};
\addlegendentry{Time Sharing $\beta=200$}

\addplot [color=blue, line width=3pt]
  table[row sep=crcr]{%
-50	0.148\\
-47.5	0.389666666666667\\
-45	0.690666666666667\\
-42.5	0.894\\
-40	0.963\\
-37.5	0.987333333333333\\
-35	0.995\\
-32.5	0.996\\
-30	0.995333333333333\\
-27.5	0.994\\
-25	0.994333333333333\\
-22.5	0.993666666666666\\
-20	0.996\\
-17.5	0.994333333333333\\
-15	0.993666666666666\\
-12.5	0.994333333333333\\
-10	0.993333333333333\\
-7.5	0.993\\
-5	0.992666666666666\\
-2.5	0.993666666666666\\
0	0.994333333333333\\
2.5	0.994666666666667\\
5	0.995\\
7.5	0.994\\
10	0.996\\
};
\addlegendentry{$\text{Concurrent }\rho\text{ = 0.4}$}

\addplot [color=green, line width=3pt]
  table[row sep=crcr]{%
-50	0.0413333333333333\\
-47.5	0.12\\
-45	0.313333333333333\\
-42.5	0.604\\
-40	0.855\\
-37.5	0.945333333333333\\
-35	0.987\\
-32.5	0.991333333333333\\
-30	0.992666666666666\\
-27.5	0.995\\
-25	0.994666666666666\\
-22.5	0.993666666666666\\
-20	0.992333333333333\\
-17.5	0.993\\
-15	0.995666666666666\\
-12.5	0.993333333333333\\
-10	0.995\\
-7.5	0.995333333333333\\
-5	0.992666666666666\\
-2.5	0.995333333333333\\
0	0.994666666666667\\
2.5	0.995\\
5	0.997\\
7.5	0.994\\
10	0.994333333333333\\
};
\addlegendentry{$\text{Concurrent }\rho\text{ = 0.8}$}

\addplot [color=red, line width=3pt]
  table[row sep=crcr]{%
-50	0.00533333333333333\\
-47.5	0.0123333333333333\\
-45	0.0343333333333333\\
-42.5	0.106\\
-40	0.298\\
-37.5	0.583666666666667\\
-35	0.813\\
-32.5	0.953333333333333\\
-30	0.983666666666666\\
-27.5	0.993\\
-25	0.994\\
-22.5	0.994666666666667\\
-20	0.996\\
-17.5	0.994666666666666\\
-15	0.994333333333333\\
-12.5	0.994666666666666\\
-10	0.995666666666666\\
-7.5	0.992333333333333\\
-5	0.993666666666666\\
-2.5	0.993\\
0	0.993666666666666\\
2.5	0.997666666666666\\
5	0.994666666666667\\
7.5	0.993\\
10	0.993666666666667\\
};
\addlegendentry{$\text{Concurrent }\rho\text{ = 1}$}

\end{axis}

\end{tikzpicture}%
}
    \caption{Sensing performance of ISAC policies against target RCS in the loose sensing requirement scenario ($T_s=5$ ms) with low-$\Delta\theta$.}
    \label{fig:PdRCS_looseSR}
\end{figure}

The probability of detection $P_D$ serves as the main sensing performance indicator, since the target must be detected within each sensing window $T_s$ to satisfy the sensing requirement.
Figs.~\ref{fig:PdRCS_strictSR} and \ref{fig:PdRCS_looseSR} show the probability of detection for the \gls{ISAC} strategies described in Section \ref{secISACstrategies} as a function of the target \gls{RCS}. Two distinct cases are considered: the \textit{strict} \gls{SR} (Fig.~\ref{fig:PdRCS_strictSR}), where the sensing window $T_s$ is small such that the average number of packets available at the \gls{BS} satisfies $\lambda_uT_s<1$, and the probability of detection of data-based \gls{ISAC} strategies doesn't approach one for sufficiently high target \gls{RCS}; and the \textit{loose} \gls{SR} (Fig.~\ref{fig:PdRCS_looseSR}), where the sensing window $T_s$ is large enough that the average number of packets available at the \gls{BS} satisfies $\lambda_uT_s>1$, and the probability of detection of all \gls{ISAC} strategies approaches one for sufficiently high target \gls{RCS}.

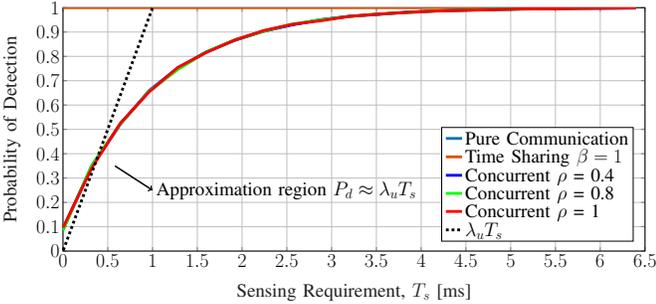
\begin{figure}[t]
\vspace{-2.2em}
    \centering
    \resizebox{0.99\columnwidth}{!}{
%
%
\definecolor{mycolor1}{rgb}{0.06600,0.44300,0.74500}%
\definecolor{mycolor2}{rgb}{0.86600,0.32900,0.00000}%
\definecolor{mycolor3}{rgb}{0.00000,1.00000,1.00000}%
\definecolor{mycolor4}{rgb}{0.12941,0.12941,0.12941}%
\begin{tikzpicture}[font=\huge]

\begin{axis}[%
width=8.521in,
height=3.566in,
scale only axis,
xmin=0,
xmax=6.5,
xlabel style={font=\color{mycolor4}, yshift=-1em},
xlabel={\huge Sensing Requirement, $T_s$ [ms]},
ymin=0,
ymax=1,
ylabel style={font=\color{mycolor4}, yshift=1em},
ylabel={\huge Probability of Detection},
axis background/.style={fill=white},
xmajorgrids,
ymajorgrids,
legend style={at={(0.98,0.04)}, anchor=south east, legend cell align=left, align=left}
]

\addplot [color=mycolor1, line width=3pt]
  table[row sep=crcr]{%
0.0016	0.0886333333333333\\
0.3216	0.350566666666667\\
0.6416	0.521066666666667\\
0.9616	0.659766666666667\\
1.2816	0.752766666666667\\
1.6016	0.818233333333333\\
1.9216	0.8679\\
2.2416	0.904466666666667\\
2.5616	0.930566666666667\\
2.8816	0.949866666666667\\
3.2016	0.964866666666667\\
3.5216	0.9735\\
3.8416	0.981\\
4.1616	0.986966666666666\\
4.4816	0.989333333333333\\
4.8016	0.992366666666667\\
5.1216	0.9947\\
5.4416	0.996\\
5.7616	0.9967\\
6.0816	0.9984\\
6.4016	0.9984\\
};
\addlegendentry{Pure Communication}

\addplot [color=mycolor2, line width=3pt]
  table[row sep=crcr]{%
0.0016	1\\
0.3216	1\\
0.6416	1\\
0.9616	1\\
1.2816	1\\
1.6016	1\\
1.9216	1\\
2.2416	1\\
2.5616	1\\
2.8816	1\\
3.2016	1\\
3.5216	1\\
3.8416	1\\
4.1616	1\\
4.4816	1\\
4.8016	1\\
5.1216	1\\
5.4416	1\\
5.7616	1\\
6.0816	1\\
6.4016	1\\
};
\addlegendentry{Time Sharing $\beta=1$}

\addplot [color=blue, line width=3pt]
  table[row sep=crcr]{%
0.0016	0.0931\\
0.3216	0.347266666666667\\
0.6416	0.5244\\
0.9616	0.6554\\
1.2816	0.7503\\
1.6016	0.818033333333334\\
1.9216	0.8662\\
2.2416	0.903033333333333\\
2.5616	0.929433333333333\\
2.8816	0.9508\\
3.2016	0.9646\\
3.5216	0.9721\\
3.8416	0.981866666666666\\
4.1616	0.986733333333333\\
4.4816	0.990466666666667\\
4.8016	0.9928\\
5.1216	0.994566666666666\\
5.4416	0.9965\\
5.7616	0.997766666666667\\
6.0816	0.9979\\
6.4016	0.998766666666666\\
};
\addlegendentry{$\text{Concurrent }\rho\text{ = 0.4}$}

\addplot [color=green, line width=3pt]
  table[row sep=crcr]{%
0.0016	0.0859\\
0.3216	0.353366666666667\\
0.6416	0.5228\\
0.9616	0.656366666666667\\
1.2816	0.745533333333333\\
1.6016	0.817766666666667\\
1.9216	0.868133333333333\\
2.2416	0.903033333333333\\
2.5616	0.932666666666667\\
2.8816	0.950366666666667\\
3.2016	0.963933333333333\\
3.5216	0.973833333333333\\
3.8416	0.980233333333334\\
4.1616	0.986333333333333\\
4.4816	0.9895\\
4.8016	0.992866666666666\\
5.1216	0.993766666666667\\
5.4416	0.996133333333333\\
5.7616	0.997633333333333\\
6.0816	0.998333333333333\\
6.4016	0.998666666666667\\
};
\addlegendentry{$\text{Concurrent }\rho\text{ = 0.8}$}

\addplot [color=red, line width=3pt]
  table[row sep=crcr]{%
0.0016	0.0969666666666667\\
0.3216	0.3415\\
0.6416	0.5258\\
0.9616	0.654\\
1.2816	0.7544\\
1.6016	0.816066666666667\\
1.9216	0.867566666666667\\
2.2416	0.9062\\
2.5616	0.932033333333333\\
2.8816	0.946366666666667\\
3.2016	0.964266666666667\\
3.5216	0.974033333333333\\
3.8416	0.980233333333333\\
4.1616	0.985933333333333\\
4.4816	0.989066666666666\\
4.8016	0.991566666666666\\
5.1216	0.9948\\
5.4416	0.9951\\
5.7616	0.997066666666667\\
6.0816	0.9979\\
6.4016	0.998666666666666\\
};
\addlegendentry{$\text{Concurrent }\rho\text{ = 1}$}

\addplot [color=black, dashed, line width=3pt] {x};
\addlegendentry{$\lambda_uT_s$}


\draw[->, ultra thick, black] (axis cs:0.58,0.35) -- (axis cs:1,0.25) 
    node [pos=1, right, font=\huge] {Approximation region $P_d\approx\lambda_uT_s$};

\end{axis}

\end{tikzpicture}%
}
    \caption{Sensing performance of ISAC policies against sensing requirement ($T_s$). The target RCS is fixed and the strong target scenario is considered.}
    \label{fig:PdSR}
\end{figure}

\begin{figure*}[t]
    \centering 
    \begin{subfigure}[t]{0.32\textwidth}
        \centering
        \resizebox{1.01\columnwidth}{!}{
\definecolor{mycolor1}{rgb}{0.06600,0.44300,0.74500}
\definecolor{mycolor2}{rgb}{0.86600,0.32900,0.00000}
\definecolor{mycolor3}{rgb}{0.12941,0.12941,0.12941}
\definecolor{mycolor4}{rgb}{0.00000, 0.70000, 0.30000}

\begin{tikzpicture}[font=\Large]
\begin{axis}[
width=5.521in,
height=3.866in,
    scale only axis,
    xmin=33,
    xmax=47,
    xlabel style={font=\color{mycolor3}},
    xlabel={\huge Communication SNR [dB]},
    ymin=0,
    ymax=1,
    ylabel style={font=\color{mycolor3}},
    ylabel={\huge Probability of Detection},
    axis background/.style={fill=white},
clip=false,
    xmajorgrids,
    ymajorgrids,
    legend style={
        at={(0.5,-0.2)}, 
        anchor=north,
        draw=none,
        fill=none,
        font=\small,
        /tikz/every even column/.append style={column sep=1cm}, 
        column sep=0.5cm,
        legend columns=3 
    }
]


\addplot [color=mycolor2, line width=3pt]
    table[row sep=crcr]{%
45.6357857104983	0\\
45.5214910926904	1\\
45.4101274315109	1\\
44.6091622915267	1\\
43.7794199408791	1\\
41.9560178575523	1\\
39.9930714061127	1\\
37.6194622481665	1\\
36.093360616105	1\\
34.9663178141921	1\\
    };

 \addplot [color=mycolor4, line width=3pt]
    table[row sep=crcr]{%
    45.6333606161049	0.3295\\
    45.4047828713926	0.332\\
    45.023767879019	0.326\\
    44.5733262635066	0.3335\\
    44.0470949732591	0.3355\\
    43.426167656579	0.331\\
    42.6768487611727	0.3395\\
    41.7382533147454	0.3405\\
    40.4858317008136	0.332\\
    38.5903841192235	0.333\\
    33.4894589563835	0.3275\\
    };

\addplot [color=mycolor1, line width=3pt, only marks, mark size=5pt, mark=o, mark options={solid, mycolor1}]
    table[row sep=crcr]{%
    45.6333606161049	0.327\\
    };


 \addplot [color=mycolor2, dashed, line width=3pt]
    table[row sep=crcr]{%
45.6357857104983	0\\
45.5214910926904	1\\
45.4101274315109	1\\
44.6091622915267	1\\
43.7794199408791	1\\
41.9560178575523	1\\
39.9930714061127	1\\
37.6194622481665	1\\
36.093360616105	1\\
34.9663178141921	1\\
    };

 \addplot [color=mycolor4, dashed, line width=3pt]
    table[row sep=crcr]{%
    45.6333606161049	0.3265\\
    45.4047843787834	0.331\\
    45.023768703007	0.3355\\
    44.5733227180624	0.334\\
    44.0470949264309	0.327\\
    43.426167656579	0.326\\
    42.6768487611727	0.334\\
    41.7382533147454	0.3315\\
    40.4858317008136	0.339\\
    38.5903633970466	0.338\\
    33.4894553572569	0.3255\\
    };

\addplot [color=mycolor1, dashed, line width=3pt, only marks, mark size=5pt, mark=square, mark options={solid, mycolor1}]
    table[row sep=crcr]{%
    45.6333606161049	0.335\\
    };
    

\addplot [color=mycolor2, dotted, line width=3pt]
    table[row sep=crcr]{%
45.6357857104983	0\\
45.5214910926904	1\\
45.4101274315109	1\\
44.6091622915267	1\\
43.7794199408791	1\\
41.9560178575523	1\\
39.9930714061127	1\\
37.6194622481665	1\\
36.093360616105	1\\
34.9663178141921	1\\
    };

 \addplot [color=mycolor4, dotted, line width=3pt]
    table[row sep=crcr]{%
    45.6333606161049	0.0005\\
    45.4047843787834	0.324\\
    45.0237654273834	0.335\\
    44.5733262635066	0.3425\\
    44.0470983935181	0.323\\
    43.426167656579	0.3375\\
    42.6768487611727	0.3305\\
    41.7382429152721	0.332\\
    40.4858173208889	0.334\\
    38.5903635300946	0.3325\\
    33.489512848213	0.3325\\
    };

\addplot [color=mycolor1, dotted, line width=3pt, only marks, mark size=6pt, mark=x, mark options={solid, mycolor1}]
    table[row sep=crcr]{%
    45.6333606161049	0.0005\\
    };


\newcommand{\legendbox}[2]{%
    \makebox[0.5cm][l]{\color{#1}\rule[0.5ex]{0.5cm}{#2pt}}%
}

\newcommand{\legenddash}[1]{%
    \tikz{\draw[#1, line width=2pt, dash pattern=on 4pt off 2pt] (0,0) -- (0.5cm,0);}%
}

\node[
    anchor=north, 
    at={(1.5in, 3.5in)}, 
    inner sep=5pt, 
    align=left,
    font=\small,
] (custom_legend) 
{
    \textbf{\Large Policy:}
    \hspace{0.5cm} \legendbox{mycolor1}{8} \Large Pure Communication
    \\[8pt]
    \hspace{2.06cm} \legendbox{mycolor2}{8} \Large Time Sharing
    \\[8pt]
    \hspace{2.06cm} \legendbox{mycolor4}{8} \Large Concurrent
    \\[8pt] 
    
   \textbf{\Large Scenario:}
    \hspace{0.5cm} \tikz{ \draw[mycolor3, solid, line width=2pt] (0,0) -- (0.5cm,0); 
    \draw[mycolor3, solid, line width=2pt, mark=o, mark options={fill=white}, mark size=5pt] plot coordinates {(0.8cm,0)};} \Large \ Low-$\mathbf{\Delta\theta}$
    \\[8pt]
    \hspace{2.6cm} \tikz{\draw[mycolor3, dashed, line width=2pt] (0,0) -- (0.5cm,0); 
    \draw[mycolor3, solid, line width=2pt, mark=square, mark options={fill=white}, mark size=5pt] plot coordinates {(0.8cm,0)};} \Large  \ Mid-$\mathbf{\Delta\theta}$
    \\[8pt]
    \hspace{2.6cm} \tikz{\draw[mycolor3, dotted, line width=2pt] (0,0) -- (0.5cm,0); 
    \draw[mycolor3, solid, line width=2pt, mark=x, mark options={fill=white}, mark size=6pt] plot coordinates {(0.8cm,0)};} \Large \ High-$\mathbf{\Delta\theta}$
};

\end{axis}
\end{tikzpicture}}
        \caption{Strict SR, strong target}\label{fig:ISACTradeoff_stricSR_strongtarget}
    \end{subfigure}
    \hfill 
    \begin{subfigure}[t]{0.32\textwidth}
        \centering
        \resizebox{1.01\columnwidth}{!}{
\definecolor{mycolor1}{rgb}{0.06600,0.44300,0.74500}
\definecolor{mycolor2}{rgb}{0.86600,0.32900,0.00000}
\definecolor{mycolor3}{rgb}{0.12941,0.12941,0.12941}
\definecolor{mycolor4}{rgb}{0.00000, 0.70000, 0.30000}

\begin{tikzpicture}[font=\Large]
\begin{axis}[
width=5.521in,
height=3.866in,
    scale only axis,
    xmin=33, 
    xmax=47, 
    xlabel style={font=\color{mycolor3}},
    xlabel={\huge Communication SNR [dB]},
    ymin=0,
    ymax=1,
    ylabel style={font=\color{mycolor3}},
    ylabel={\huge Probability of Detection},
    axis background/.style={fill=white},
    xmajorgrids,
    ymajorgrids,
    legend style={
        at={(0.5,-0.2)}, 
        anchor=north,
        draw=none,
        fill=none,
        font=\small,
        /tikz/every even column/.append style={column sep=1cm}, 
        column sep=0.5cm,
        legend columns=3 
    }
]


 \addplot [color=mycolor2, line width=3pt]
    table[row sep=crcr]{%
45.6357857104983	0\\
45.5214910926904	0.009875\\
45.4101274315109	0.02875\\
44.6091622915267	0.39875\\
43.7794199408791	0.736375\\
41.9560178575523	0.96075\\
39.9930714061127	0.989\\
37.6194622481665	0.998\\
36.093360616105	0.99675\\
34.9663178141921	0.99825\\
    };

 \addplot [color=mycolor4, line width=3pt]
    table[row sep=crcr]{%
45.6333606161049	0.3315\\
45.4047810639651	0.327\\
45.0237678930341	0.325\\
44.5733223804009	0.324\\
44.0470996951053	0.3365\\
43.426167656579	0.342\\
42.6768487611727	0.3405\\
41.7382533147454	0.327\\
40.4858317008136	0.3335\\
38.5903906746209	0.326\\
33.4894574973738	0.327\\
    };

\addplot [color=mycolor1, line width=3pt, only marks, mark size=5pt, mark=o, mark options={solid, mycolor1}]
    table[row sep=crcr]{%
45.6333606161049	0.337\\
    };


\addplot [color=mycolor2, dashed, line width=3pt]
    table[row sep=crcr]{%
45.6357857104983	0\\
45.5214910926904	0.00375\\
45.4101274315109	0.008\\
44.6091622915267	0.192625\\
43.7794199408791	0.497125\\
41.9560178575523	0.861875\\
39.9930714061127	0.95425\\
37.6194622481665	0.98325\\
36.093360616105	0.989375\\
34.9663178141921	0.990625\\
    };

\addplot [color=mycolor4, dashed, line width=3pt]
    table[row sep=crcr]{%
45.6333606161049	0.0565\\
45.4047827807512	0.2485\\
45.0237650665807	0.28\\
44.5733262635066	0.279\\
44.0470947802023	0.296\\
43.4261615582791	0.309\\
42.6768487611727	0.3115\\
41.7382533147454	0.344\\
40.4858317008136	0.327\\
38.590362561192	0.325\\
33.4894571201936	0.324\\
    };

\addplot [color=mycolor1, dashed, line width=3pt, only marks, mark size=5pt, mark=square, mark options={solid, mycolor1}]
    table[row sep=crcr]{%
45.6333606161049	0.0655\\
    };
    

 \addplot [color=mycolor2, dotted, line width=3pt]
    table[row sep=crcr]{%
45.6357857104983	0\\
45.5214910926904	0.00825\\
45.4101274315109	0.027625\\
44.6091622915267	0.25175\\
43.7794199408791	0.451625\\
41.9560178575523	0.675125\\
39.9930714061127	0.75625\\
37.6194622481665	0.82375\\
36.093360616105	0.833375\\
34.9663178141921	0.840875\\
    };

 \addplot [color=mycolor4, dotted, line width=3pt]
    table[row sep=crcr]{%
45.6333606161049	0.0005\\
45.4047843787834	0.0855\\
45.0237652643651	0.2115\\
44.5733262071258	0.2475\\
44.0470952633881	0.267\\
43.4261610816799	0.281\\
42.6768408935771	0.2975\\
41.7382482289354	0.319\\
40.4858317008136	0.322\\
38.5903613358655	0.326\\
33.4894600011426	0.3155\\
    };

\addplot [color=mycolor1, dotted, line width=3pt, only marks, mark size=6pt, mark=x, mark options={solid, mycolor1}]
    table[row sep=crcr]{%
45.6333606161049	0.001\\
    };


\newcommand{\legendbox}[2]{%
    \makebox[0.5cm][l]{\color{#1}\rule[0.5ex]{0.5cm}{#2pt}}%
}

\newcommand{\legenddash}[1]{%
    \tikz{\draw[#1, line width=2pt, dash pattern=on 4pt off 2pt] (0,0) -- (0.5cm,0);}%
}

\node[
    anchor=north, 
    at={(1.5in, 2.9in)}, 
    inner sep=5pt, 
    align=left,
    font=\small,
] (custom_legend) 
{
    \textbf{\Large Policy:}
    \hspace{0.5cm} \legendbox{mycolor1}{8} \Large Pure Communication
    \\[8pt]
    \hspace{2.06cm} \legendbox{mycolor2}{8} \Large Time Sharing
    \\[8pt]
    \hspace{2.06cm} \legendbox{mycolor4}{8} \Large Concurrent
    \\[8pt] 
    
   \textbf{\Large Scenario:}
    \hspace{0.5cm} \tikz{ \draw[mycolor3, solid, line width=2pt] (0,0) -- (0.5cm,0); 
    \draw[mycolor3, solid, line width=2pt, mark=o, mark options={fill=white}, mark size=5pt] plot coordinates {(0.8cm,0)};} \Large \ Low-$\mathbf{\Delta\theta}$
    \\[8pt]
    \hspace{2.6cm} \tikz{\draw[mycolor3, dashed, line width=2pt] (0,0) -- (0.5cm,0); 
    \draw[mycolor3, solid, line width=2pt, mark=square, mark options={fill=white}, mark size=5pt] plot coordinates {(0.8cm,0)};} \Large  \ Mid-$\mathbf{\Delta\theta}$
    \\[8pt]
    \hspace{2.6cm} \tikz{\draw[mycolor3, dotted, line width=2pt] (0,0) -- (0.5cm,0); 
    \draw[mycolor3, solid, line width=2pt, mark=x, mark options={fill=white}, mark size=6pt] plot coordinates {(0.8cm,0)};} \Large \ High-$\mathbf{\Delta\theta}$
};

\end{axis}
\end{tikzpicture}}
        \caption{Strict SR, weak target}\label{fig:ISACTradeoff_stricSR_weaktarget}
    \end{subfigure}
    \hfill 
    \begin{subfigure}[t]{0.32\textwidth}
        \centering
        \resizebox{1.01\columnwidth}{!}{
\definecolor{mycolor1}{rgb}{0.06600,0.44300,0.74500}
\definecolor{mycolor2}{rgb}{0.86600,0.32900,0.00000}
\definecolor{mycolor3}{rgb}{0.12941,0.12941,0.12941}
\definecolor{mycolor4}{rgb}{0.00000, 0.70000, 0.30000}

\begin{tikzpicture}[font=\Large]
\begin{axis}[
width=5.521in,
height=3.866in,
    scale only axis,
    xmin=33, 
    xmax=47, 
    xlabel style={font=\color{mycolor3}},
    xlabel={\huge Communication SNR [dB]},
    ymin=0,
    ymax=1,
    ylabel style={font=\color{mycolor3}},
    ylabel={\huge Probability of Detection},
    axis background/.style={fill=white},
    xmajorgrids,
    ymajorgrids,
    legend style={
        at={(0.5,-0.2)}, 
        anchor=north,
        draw=none,
        fill=none,
        font=\small,
        /tikz/every even column/.append style={column sep=1cm}, 
        column sep=0.5cm,
        legend columns=3 
    }
]


 \addplot [color=mycolor4, line width=3pt]
    table[row sep=crcr]{%
45.6333606161049	0.993958333333333\\
45.4047837167568	0.992916666666667\\
45.0237671351246	0.995\\
44.5733250995952	0.994791666666667\\
44.0470986921434	0.993333333333333\\
43.4261658730992	0.993333333333334\\
42.6768461923666	0.994375\\
41.7382500184147	0.992083333333333\\
40.485825643095	0.993958333333333\\
38.590377178639	0.995416666666667\\
33.4894897098802	0.99375\\
    };

\addplot [color=mycolor1, line width=3pt, only marks, mark size=5pt, mark=o, mark options={solid, mycolor1}]
    table[row sep=crcr]{%
45.6333606161049	0.994791666666667\\
    };

 \addplot [color=mycolor2, line width=3pt]
    table[row sep=crcr]{%
45.6357857104983	0\\
45.6288425518346	0.00833333333333334\\
45.6219104756317	0.0245833333333333\\
45.5668486310192	0.547291666666667\\
45.4989887375863	0.931875\\
45.3015481556287	0.999791666666667\\
44.9912058182291	1\\
44.4300463984397	1\\
43.9331685565486	1\\
43.4873472300212	1\\
    };


\addplot [color=mycolor4, dashed, line width=3pt]
    table[row sep=crcr]{%
45.6333606161049	0.902708333333333\\
45.4047837167568	0.989791666666667\\
45.0237671351246	0.994791666666667\\
44.5733250995952	0.994375\\
44.0470986921434	0.993333333333333\\
43.4261658730992	0.993333333333334\\
42.6768461923666	0.994375\\
41.7382500184147	0.991875\\
40.485825643095	0.99375\\
38.590377178639	0.995416666666667\\
33.4894897098802	0.99375\\
    };

\addplot [color=mycolor1, dashed, line width=3pt, only marks, mark size=5pt, mark=square, mark options={solid, mycolor1}]
    table[row sep=crcr]{%
45.6333606161049	0.904166666666667\\
    };

\addplot [color=mycolor2, dashed, line width=3pt]
    table[row sep=crcr]{%
45.6357857104983	0\\
45.6288425518346	0.004375\\
45.6219104756317	0.010625\\
45.5668486310192	0.304166666666667\\
45.4989887375863	0.792708333333333\\
45.3015481556287	0.999791666666667\\
44.9912058182291	1\\
44.4300463984397	1\\
43.9331685565486	1\\
43.4873472300212	1\\
    };
    

 \addplot [color=mycolor4, dotted, line width=3pt]
    table[row sep=crcr]{%
45.6333606161049	0.00270833333333333\\
45.4047837167568	0.871458333333333\\
45.0237671351246	0.985208333333333\\
44.5733250995952	0.99\\
44.0470986921434	0.991666666666667\\
43.4261658730992	0.992083333333333\\
42.6768461923666	0.993333333333333\\
41.7382500184147	0.991875\\
40.485825643095	0.993333333333333\\
38.590377178639	0.995208333333333\\
33.4894897098802	0.99375\\
    };

\addplot [color=mycolor1, dotted, line width=3pt, only marks, mark size=6pt, mark=x, mark options={solid, mycolor1}]
    table[row sep=crcr]{%
45.6333606161049	0.00333333333333333\\
    };

 \addplot [color=mycolor2, dotted, line width=3pt]
    table[row sep=crcr]{%
45.6357857104983	0\\
45.6288425518346	0.0114583333333333\\
45.6219104756317	0.0289583333333333\\
45.5668486310192	0.336875\\
45.4989887375863	0.630208333333333\\
45.3015481556287	0.910833333333333\\
44.9912058182291	0.984583333333333\\
44.4300463984397	0.998541666666667\\
43.9331685565486	1\\
43.4873472300212	1\\
    };

\newcommand{\legendbox}[2]{%
    \makebox[0.5cm][l]{\color{#1}\rule[0.5ex]{0.5cm}{#2pt}}%
}

\newcommand{\legenddash}[1]{%
    \tikz{\draw[#1, line width=2pt, dash pattern=on 4pt off 2pt] (0,0) -- (0.5cm,0);}%
}

\node[
    anchor=north, 
    at={(1.5in, 2.9in)}, 
    inner sep=5pt, 
    align=left,
    font=\small,
] (custom_legend) 
{
    \textbf{\Large Policy:}
    \hspace{0.5cm} \legendbox{mycolor1}{8} \Large Pure Communication
    \\[8pt]
    \hspace{2.06cm} \legendbox{mycolor2}{8} \Large Time Sharing
    \\[8pt]
    \hspace{2.06cm} \legendbox{mycolor4}{8} \Large Concurrent
    \\[8pt] 
    
   \textbf{\Large Scenario:}
    \hspace{0.5cm} \tikz{ \draw[mycolor3, solid, line width=2pt] (0,0) -- (0.5cm,0); 
    \draw[mycolor3, solid, line width=2pt, mark=o, mark options={fill=white}, mark size=5pt] plot coordinates {(0.8cm,0)};} \Large \ Low-$\mathbf{\Delta\theta}$
    \\[8pt]
    \hspace{2.6cm} \tikz{\draw[mycolor3, dashed, line width=2pt] (0,0) -- (0.5cm,0); 
    \draw[mycolor3, solid, line width=2pt, mark=square, mark options={fill=white}, mark size=5pt] plot coordinates {(0.8cm,0)};} \Large  \ Mid-$\mathbf{\Delta\theta}$
    \\[8pt]
    \hspace{2.6cm} \tikz{\draw[mycolor3, dotted, line width=2pt] (0,0) -- (0.5cm,0); 
    \draw[mycolor3, solid, line width=2pt, mark=x, mark options={fill=white}, mark size=6pt] plot coordinates {(0.8cm,0)};} \Large \ High-$\mathbf{\Delta\theta}$
};

\end{axis}
\end{tikzpicture}}
        \caption{Loose SR, weak target}\label{fig:ISACTradeoff_looseSR_weaktarget}
    \end{subfigure}
    
    \caption{\gls{ISAC} trade-off curves in different scenarios obtained for different values of $\rho\in[0,1]$ and $\beta\in[0,200]$ for \textit{concurrent transmission} and \textit{time sharing} policies, respectively.}\label{fig:TradeoffCurves}
\end{figure*}
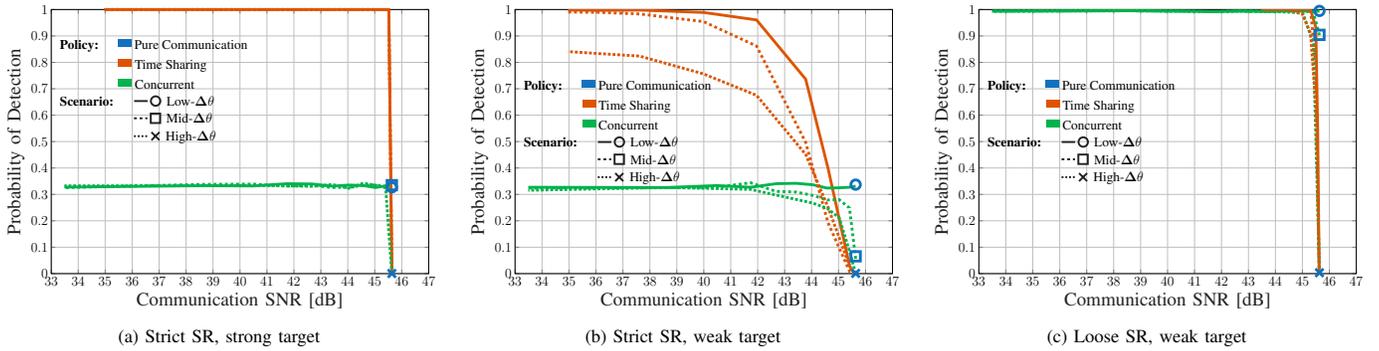
In these scenarios, the following effects are identified:
\begin{itemize}
    \item \textit{C-SNR-Boost}: as \textit{concurrent transmission} and \textit{pure communication} \gls{ISAC} strategies exploit data symbols for sensing \cite{keskin2025fundamental}, a larger number of observations can be collected within a sensing window compared to the \textit{time sharing} strategy, which relies on dedicated pilot signals. This leads to an effective \gls{SNR} boost and improves detection capabilities with respect to \textit{time sharing} \cite{keskin2025fundamental}.

    \item \textit{NFB-Loss (non-full buffer loss)}: as data are not always available at the \gls{BS} due to the bursty traffic assumption, under a strict sensing requirement and \textit{concurrent transmission} or \textit{pure communication} policies, the \gls{BS} may not collect any observation within the sensing window $T_s$. As a result, the \gls{BS} is not able to detect the target and the probability of detection drops. This effect places \textit{time sharing} at an advantage over data-based strategies.

    \item \textit{DM-Loss (directional masking loss)}:  as the relative position between the target and the \gls{UE} impacts sensing performance, if the angular deviation between them becomes large, the \textit{pure communication} policy may no longer detect the target (i.e., subspace trade-off \cite{Liu2023}).
\end{itemize}

In particular, Fig.~\ref{fig:PdRCSsmallSRsmalltheta} shows that for weak targets, the C-SNR-boost effect allows \textit{pure communication} and \textit{concurrent transmission} to achieve higher sensing performance compared to \textit{time sharing} with $\beta=1$ and $\beta=200$. As expected, as $\beta$ is increased, the C-SNR-boost effect becomes less visible, at the cost of deteriorating the communication \gls{SNR}. On the other hand, for stronger targets, \textit{time sharing} achieves a probability of detection of one, whereas data-based \gls{ISAC} policies achieve low probability of detection due to NFB-loss effect induced by the stringent \gls{SR}. Moreover, in Fig.~\ref{fig:PdRCSmidSRmidtheta}, it can be noticed that, as the angular separation between the user and the target is increased in the mid-$\Delta\theta$ scenario, the DM-loss makes \textit{pure communication} to achieve lower detection capabilities than \textit{concurrent transmission} for weak targets. Conversely, for stronger targets the probability of detection of the \textit{pure communication} strategy increases, reaching the NFB-loss induced saturation level as for the \textit{concurrent transmission} policy. When considering the high-$\Delta\theta$ case study shown in Fig.~\ref{fig:PdRCSbigSRbigtheta}, the DM-loss effect, as expected, renders the \textit{pure communication} policy inadequate for sensing. Finally, when the sensing requirement is relaxed, as shown in Fig.~\ref{fig:PdRCS_looseSR}, the NFB-loss becomes negligible, and the C-SNR-boost effect makes data-based strategies to achieve better sensing performance for weak targets.

The impact of the sensing requirement is explicitly shown in Fig.~\ref{fig:PdSR}, where the probability of detection is shown for different values of $T_s$. The low-$\Delta\theta$ scenario with a strong target is considered to illustrate that, as the sensing requirement is relaxed, the probability of detection under the \textit{pure communication} and \textit{concurrent transmission} \gls{ISAC} strategies approaches one. Similar conclusions can be drawn for other scenarios as well. Specifically, Fig.~\ref{fig:PdSR} shows that the probability of detection begins to approach one for $T_s \approx 4$ ms. Therefore, when $T_s < 4$ ms, the NFB-loss causes the probability of detection to saturate. Conversely, when $T_s > 4$ ms, the probability of detection approaches one, as the \gls{GLRT}-based detector with coherent integration collects more observations. Moreover, Figs.~\ref{fig:PdRCS_strictSR} and \ref{fig:PdSR} indicate that, under a stringent sensing requirement, the probability of detection for data-based \gls{ISAC} policies is approximately proportional to $\lambda_uT_s$, which represents the average number of packets available at the \gls{BS} within a sensing window. Intuitively, this approximation holds since the \gls{BS} can detect a target only when communication packets are available, and cannot collect any observations otherwise.

\subsection{ISAC Trade-offs Curves and Design Guidelines}
In order to investigate \gls{ISAC} trade-offs, Fig.~\ref{fig:TradeoffCurves} shows the communication and sensing performance for all the \gls{ISAC} strategies described in Section \ref{secISACstrategies}. In particular, the probability of detection and the communication \gls{SNR} are shown as performance metrics for the sensing and communication functionalities, respectively.
Each curve is obtained by sweeping trade-off parameters as $\rho\in[0,1]$ for \textit{concurrent transmission} and $\beta\in[0,200]$ for \textit{time sharing}.
The results are presented for both strong-target (the target RCS is large, $\sigma_{rcs}=5$ dBsm) and weak-target (the target RCS is small, $\sigma_{rcs}=-20$ dBsm) cases, respectively in Fig.~\ref{fig:ISACTradeoff_stricSR_strongtarget} and Figs.~\ref{fig:ISACTradeoff_stricSR_weaktarget} and \ref{fig:ISACTradeoff_looseSR_weaktarget}.
Moreover, the impact of geometry on \gls{ISAC} trade-offs is shown in Fig.~\ref{fig:TradeoffCurves} by including the varying $\Delta\theta$ scenarios described in Section \ref{secSimScenario}.

From Fig.~\ref{fig:ISACTradeoff_stricSR_strongtarget}, it can be observed that, when the sensing requirement is stringent and the communication functionality demands maximum communication \gls{SNR}, the \textit{time sharing} policy with $\beta=1$ achieves near-optimal \gls{ISAC} performance in the strong-target case. Conversely, Fig.~\ref{fig:ISACTradeoff_stricSR_weaktarget} shows that, when the target is weak, (near) maximum communication \gls{SNR} can be achieved by adopting either the \textit{pure communication} or \textit{concurrent transmission} (with low $\rho$) policies, at the cost of a lower probability of detection. In this case, the C-SNR-boost effect enables data-based \gls{ISAC} policies to achieve better sensing performance than \textit{time sharing} (with low $\beta$). At the same time, due to the strict sensing requirement, the NFB-loss limits the achievable $P_D$ for data-only policies. Therefore, under strict \gls{SR} and weak-target conditions, maximum communication performance can be obtained only with limited sensing performance.
On the other hand, Fig.~\ref{fig:ISACTradeoff_stricSR_weaktarget} reveals also that, if the communication requirement is relaxed, the \gls{BS} can adopt \textit{time sharing} with higher $\beta$ to obtain better sensing performance.
If the sensing requirement is less stringent, \textit{time sharing} enables optimal scenario-independent \gls{ISAC} performance in the strong-target case. Conversely, when the target is weak, the \textit{pure communication} and \textit{concurrent transmission} strategies achieve near-maximum communication \gls{SNR} in the low-$\Delta\theta$ and mid/high-$\Delta\theta$ regimes, respectively, as depicted in Fig.~\ref{fig:ISACTradeoff_looseSR_weaktarget}.
Finally, when both communication and sensing requirements are relaxed, the \textit{concurrent transmission} policy (with sufficiently high $\rho$) achieves optimal \gls{ISAC} performance regardless of the relative position between the target and the user, as well as the target strength.

Guidelines for selecting the most suitable \gls{ISAC} strategy in different scenarios under bursty traffic, considering the effects discussed in Section \ref{sec:PdvsRCSandEffects} and the results shown in Fig.~\ref{fig:TradeoffCurves}, are provided in Table \ref{tableGuidelines}.

\begin{table}[t]
\centering
\caption{Guidelines on which scenarios are best suited for different ISAC policies}\label{TabGuide}
{
\begin{tabular}{|| >{\centering}p{2cm} | c | c | c ||}
\hline
\diagbox[width=2.43cm]{\makecell[l]{ISAC \\ Requirements}}{\makecell{ISAC \\ Policy}} & \makecell{Pure \\ communication} & \makecell{Concurrent \\ transmission} & \makecell{Time \\ sharing} \\
\hline
\makecell{Strict SR \\ Maximum $\text{SNR}_c$} & \makecell{Low-$\Delta\theta$ \\ weak target \\ (Low $P_D$)} & \makecell{Mid/high-$\Delta\theta$ \\ weak target \\ (Low $P_D$)} & \makecell{Strong target \\ (High $P_D$)} \\
\hline
\makecell{Loose SR \\ Maximum $\text{SNR}_c$} & \makecell{Low-$\Delta\theta$ \\ weak target \\ (High $P_D$)} & \makecell{Mid/high-$\Delta\theta$ \\ weak target \\ (High $P_D$)} & \makecell{Strong target \\ (High $P_D$)} \\
\hline 
\makecell{Strict SR \\ Low $\text{SNR}_c$} & $-$ & $-$ & \makecell{Strong target \\ Weak target\\ (High $P_D$)} \\
\hline
\makecell{Loose SR \\ Low $\text{SNR}_c$} & $-$ & \makecell{Strong target \\ Weak target \\ (High $P_D$)} & $-$ \\
\hline 
\end{tabular}\label{tableGuidelines}
}
\end{table}

\section{Conclusions and Future Works}
In this work, spatial trade-offs in monostatic \gls{ISAC} are revised by incorporating a bursty traffic model. Specifically, three policies are considered: \textit{pure communication}, which prioritizes communication and treats sensing as opportunistic; \textit{concurrent transmission}, which relies solely on random data and balances \gls{ISAC} performance by allocating power to sensing and communication beams; and finally, \textit{time sharing}, which relies on dedicated pilots for sensing.
Moreover, a \gls{GLRT}-based sensing algorithm is proposed for multi-target detection by integrating across multiple observations collected within the sensing window. Numerical results reveal that different effects arise depending on the traffic conditions, the environment's geometry, data utilization, and \gls{ISAC} requirements.
Finally, scenario-dependent guidelines for the efficient design of the transmit precoder under distinct \gls{ISAC} requirements are provided. In future research, this investigation of spatial trade-offs under bursty traffic conditions will be extended to multi-user, multi-target, and multi-carrier transmission, also including multipath and Doppler effects.

\balance

\bibliographystyle{IEEEtran}
\bibliography{references}

\end{document}